\documentstyle[twoside,fleqn,espcrc2,epsf]{article}

\newcommand{\be}{\begin{equation}}
\newcommand{\ee}{\end{equation}}
\newcommand{\bea}{\begin{eqnarray}}
\newcommand{\eea}{\end{eqnarray}}

\newcommand{\ov}{\overline}
\newcommand{\bvec}{\mathbf}

\newcommand{\gapproxeq}
{\lower .7ex\hbox{$\;\stackrel{\textstyle >}{\sim}\;$}}
\newcommand{\lapproxeq}
{\lower .7ex\hbox{$\;\stackrel{\textstyle <}{\sim}\;$}}

\newcommand{\hh}{\hat{h}}
\newcommand{\hH}{\widehat{H}}
\newcommand{\hmu}{\widehat{M}_u}
\newcommand{\hdmj}{\Delta \widehat{M}_j}
\newcommand{\hgi}{\widehat{\Gamma}_i}
\newcommand{\hgj}{\widehat{\Gamma}_j}
\newcommand{\hrho}{\hat{\rho}}
\newcommand{\hp}{\hat{\mathrm p}}

\def\neb{\hbox{$\ov{\nu}_e \!$ }}

\def\ca{{C_{\scriptscriptstyle A}}}
\def\cv{{C_{\scriptscriptstyle V}}}

\newcommand{\tre}{\left( \cv^2 + 3 \ca^2 \right)}

\newcommand{\ApJ}{{\it Astrophys. J. \,}}
\newcommand{\ApJS}{{\it Astrophys. J. Suppl. \,}}

\newcommand{\NP}{{\it Nucl. Phys. \,}}
\newcommand{\PR}{{\it Phys. Rev. \,}}

\newcommand{\etal}{{\it et al.}}

\newcommand{\AmS}{{\protect\the\textfont2
  A\kern-.1667em\lower.5ex\hbox{M}\kern-.125emS}}

\title{\hfill $\mbox{\small{
$\stackrel{\rm\textstyle DSF-29/99\quad}
{\rm\textstyle astro-ph/9907420 \quad}$}}$ \\[.5truecm]
Primordial Nucleosynthesis: Accurate Predictions}

\author{S. Esposito, G. Mangano, G. Miele, and
O. Pisanti\address{Dipartimento di Fisica, Universit\'{a} di Napoli
"Federico II", and INFN,
Sezione di Napoli, Mostra D'Oltremare Pad. 20, I-80125 Napoli, Italy}%
\thanks{Talk given by G. Miele at the {\it International Workshop on 
Particles in Astrophysics and Cosmology: From Theory to Observation}, 
Valencia 1999.}
}

\begin{document}


\maketitle

\begin{abstract}
A new accurate evaluation of primordial light nuclei abundances is
presented. The proton to neutron conversion rates have been corrected to
take into account radiative effects, finite nucleon mass, thermal and
plasma corrections. The theoretical uncertainty on $^4He$ is so reduced to
the order of 0.1 \%.
\end{abstract}

Big Bang Nucleosynthesis (BBN) represents a  key subject of modern
cosmology since it is a powerful tool to study fundamental interactions. In
the recent years, the improvement on the measurement accuracy of light
primordial nuclei abundances allowed BBN to enter in a sort of {\it
precision} era. In view of this, a great theoretical effort has been
devoted to make theoretical predictions comparably accurate. Unfortunately,
as far as $D$, $^3He$ and $^7Li$ are concerned, the theoretical uncertainty
on their primordial abundances is greatly dominated by a poor knowledge of
many nuclear reactions involved in their production. On the contrary, the
$^4He$ abundance results into a robust prediction and thus an effort to
reduce at less than $1 \%$ its theoretical uncertainty is meaningful.

To improve the accuracy on the prediction of $^4He$ abundance in Ref.
\cite{EMMP1} we performed a thoroughly analysis of all corrections to the
proton/neutron conversion rates, $
\nu_e + n \leftrightarrow e^- + p$,
$e^+ + n \leftrightarrow \neb + p$, $n \leftrightarrow e^- + \neb + p$
which fix at the freeze out temperature $\sim 1~MeV$ the neutron to proton
density ratio. The Born rates, obtained in the tree level $V-A$ limit and
with infinite nucleon mass, have been corrected to take into account three
classes of relevant effects: electromagnetic radiative corrections, finite
nucleon mass corrections and plasma effects.

\section{The total corrections to Born rates}

Let us consider for example, the averaged rate per nucleon for process $n
\rightarrow e^- + \neb + p$. In the simple $V-A$ tree level, and in the
limit of infinite nucleon mass ({\it Born approximation}), one has
\bea
\omega_B = \frac{G_F^2 \tre}{2
\pi^3} \, \int_0^\infty d {|\bvec{p}'| \,|\bvec{p}'|^2} \,  q_0^2
\nonumber\\
{\times} \, \Theta(q_0) \,\left[ 1 - F_\nu (q_0)\right]  \left[ 1 - F_e(p_0')
\right],~~~~
\label{neutron}
\eea
$\bvec{p}'$ and $p_0'$ are the electron momentum and energy, and $q_0$ the
neutrino energy. The functions $F_\nu$ and $F_e$ denote the neutrino
(antineutrino) and electron (positron) Fermi distributions, respectively
\cite{EMMP1,EMMP2}.

The accuracy of Born approximation can be tested by comparing, for example,
the prediction for neutron lifetime obtained from (\ref{neutron}) in the
limit of vanishing temperature, $\tau_n \simeq 961~s$, with the
experimental value $\tau_n^{ex} = (886.7 {\pm} 1.9)~s$ \cite{PDG98}. To recover
the experimental value, a correction of about $8\%$ is expected to come
from radiative and/or finite nucleon mass effects. In Figure 1 the Born
rates for $n \rightarrow p$ processes are reported. In Figure 2 the total
corrections, listed above, to Born rates are shown. They are essentially
dominated by radiative and {\it kinetic} corrections which at low
temperature amount to $8\%$ of the total rates.

\begin{figure}[htb]
\vspace{-1cm}
\epsfysize=7.0cm
\epsfxsize=7.0cm
\centerline{\epsffile{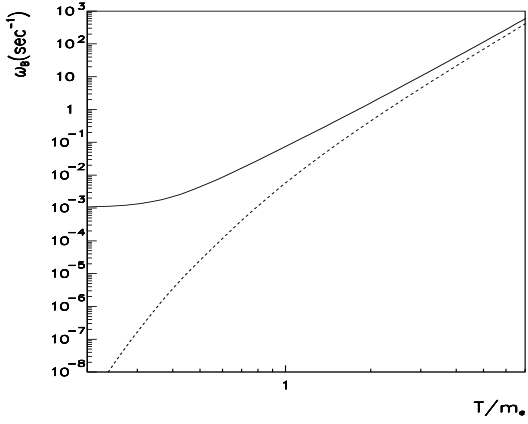}}
\vspace{-1.2truecm}
\caption{The total Born rates, $\omega_B$, for $n \rightarrow p$ (solid
line) and $p \rightarrow n$ transitions (dashed line).}
\end{figure}

\begin{figure}[htb]
\vspace{-1cm}
\epsfysize=7.0cm
\epsfxsize=7.0cm
\centerline{\epsffile{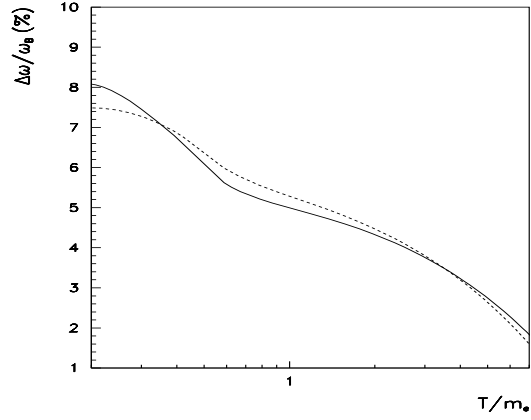}}
\vspace{-1.2truecm}
\caption{The total corrections to Born rates for $n \leftrightarrow p$
transitions (the same notation of Figure 1 is adopted).}
\end{figure}

\section{The set of equations for BBN }

The BBN equations \cite{EMMP2} can be transformed in a set of $N_{nuc}+1$
differential equations for $\hh \equiv n_B/T^3$ and the nuclide relative
abundances $X_i$ with $z=m_e/T$ as the evolution parameter. In terms of
these new variables the BBN set of equations becomes
\be
\frac{d \hh}{dz} \, = \, \left[ 1-\hH~G\right] \frac{3\hh}{z}
,
\label{e:basic1}
\ee
\be
\frac{dX_i}{dz} \, = \, G~ \frac{\hgi}{z} ~~~ ~~~~~~ i = 1, ... , N_{nuc}
~~~,
\label{e:basic2}
\ee
where $(\Theta = \Theta(z_D-z))$
\bea
G=\left[ \sum_\alpha (4 \hrho_\alpha - z \frac{\partial
\hrho_\alpha}{\partial z}) + 4 \Theta \hrho_\nu + \frac 32 \hh
\sum_j X_j\right]\nonumber\\
{\times}
\left\{3 \left[ \sum_\alpha (\hrho_\alpha + \hp_\alpha) + \frac 43
\Theta \hrho_\nu + \hh \sum_j X_j \right] \hH \right. \nonumber\\
+ \left. \hh \sum_j \left( z \hdmj + \frac 32 \right) \hgj  \right\}^{-1}
\eea
In the previous equations $z_D=m_e/T_D$ ($T_D$ is the neutrino decoupling
temperature), $\alpha=e,\gamma$, and the dimensionless Hubble parameter
$\hH=H/m_e$ reads
\bea
\hH= \sqrt{\frac{8 \pi}{3}} \frac{m_e}{M_{P}} \frac{1}{z^2}
\left[\hrho_\gamma + \hrho_e + \hrho_\nu \right. \nonumber\\
+ \left.\hh \left( z \hmu + \sum_j
\left( z \hdmj + \frac 32 \right) X_j \right) \right]^{1/2}.
\label{e:hatHubble}
\eea
The quantities $\hmu =M_u/m_e$, $\hdmj = \Delta M_j/m_e$, $\hgj =
\Gamma_j/m_e$ are the dimensionless atomic mass unit, mass excess and rate,
respectively.

The initial value for
(\ref{e:basic1}) is provided in terms of the final
baryon to photon density ratio $\eta$ according to the equation
\be
\hh_{in} = \frac{2 \zeta(3)}{\pi^2} \eta_{in} = \frac{11}{4}~ \frac{2
\zeta(3)}{\pi^2} \eta~~~.
\ee
The condition of Nuclear Statistical Equilibrium (NSE), very well satisfied
at the initial temperature $T_{in}=10 ~MeV$, fixes the initial nuclide
relative abundances. From NSE one gets
\bea
X_i(T_{in}) = \frac{g_i}{2} \left( \zeta(3)\sqrt{\frac{8}{ \pi}}
\right)^{A_i-1} A_i^{\frac{3}{2}} \eta^{A_i-1}\nonumber\\
{\times}
\left( \frac{T_{in}}{M_N}
\right)^{\frac{3}{2} (A_i-1)} \, X_p^{Z_i} \, X_n^{A_i-Z_i} \,
e^{\frac{B_i}{T_{in}}},
\label{e:s9}
\eea
where $B_i$ denotes the binding energy.

\section{Light Element Abundances}

By using in the BBN equations the corrected rates for $n \leftrightarrow p$
processes one can predict, with high accuracy, the primordial values for
$D$, $^3He$, $^4He$ and $^7Li$
\bea
Y_2  = \frac{X_3}{X_2}~,~~~~~~~~~~~~Y_3 =
\frac{X_5}{X_2}~,\nonumber\\ Y_4 =
\frac{M_6~ X_6}{\sum_j M_j~ X_j} ~, ~~~~~ Y_7 = \frac{X_8}{X_2} ~~~.
\eea
Using the results of \cite{Fiorentini} to quantify the uncertainties coming
from nuclear reaction processes, on can observe that only for $Y_4$ the
correction to Born rates affects result on $Y_4$ by an amount larger than
the theoretical uncertainties, including nuclear reactions. For $D$, $^3He$
and $^7Li$ the uncertainty, due to the poor knowledge of nuclear reaction
rates, is estimated to be of the order of $(10 \div 30)\%$
\cite{Fiorentini}, thus much less than the effects of radiative/thermal
correction on $n \leftrightarrow p$ rates.

In Fig. 3 the predictions on $Y_4$ are shown versus $\eta$ for
$N_\nu=2,3,4$ and for a $1~\sigma$ variation of $\tau_n^{ex}$. The two
experimental estimates for the primordial $^4He$ mass fraction,
$Y_4^{(l)}=0.234 {\pm} 0.002 {\pm} 0.005$ and $Y_4^{(h)}=0.243 {\pm} 0.003$ (see for
example \cite{Sarkar}) are the horizontal bands. Fig.s 4 and 5 show the
predictions for $D$ and $^7Li$ abundances. Note that, due to the negligible
variation of $Y_2$ and $Y_7$ on small $\tau_n$ changes, no splitting of
predictions for $1~\sigma$ variation of $\tau_n^{ex}$ is present.

\begin{figure}[htb]
\vspace{-1cm}
\epsfysize=7.0cm
\epsfxsize=7.0cm
\centerline{\epsffile{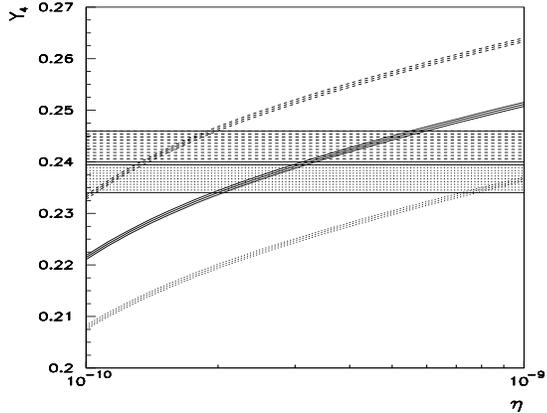}}
\vspace{-1.2truecm}
\caption{The $^4He$ mass fraction, $Y_4$, versus $\eta$. The three
solid lines are, from larger to lower values of $Y_4$, the predictions
corresponding to $N_\nu=3$ and $\tau_n^{ex}=888.6~s$, $886.7~s$, $884.8~s$,
respectively. Analogously, the dashed lines correspond to $N_\nu=4$ and the
dotted ones to $N_\nu=2$. The dotted and dashed horizontal band are the
experimental values $Y_4^{(l)}$ and $Y_4^{(h)}$}
\end{figure}

\begin{figure}[htb]
\vspace{-1cm}
\epsfysize=7.0cm
\epsfxsize=7.0cm
\centerline{\epsffile{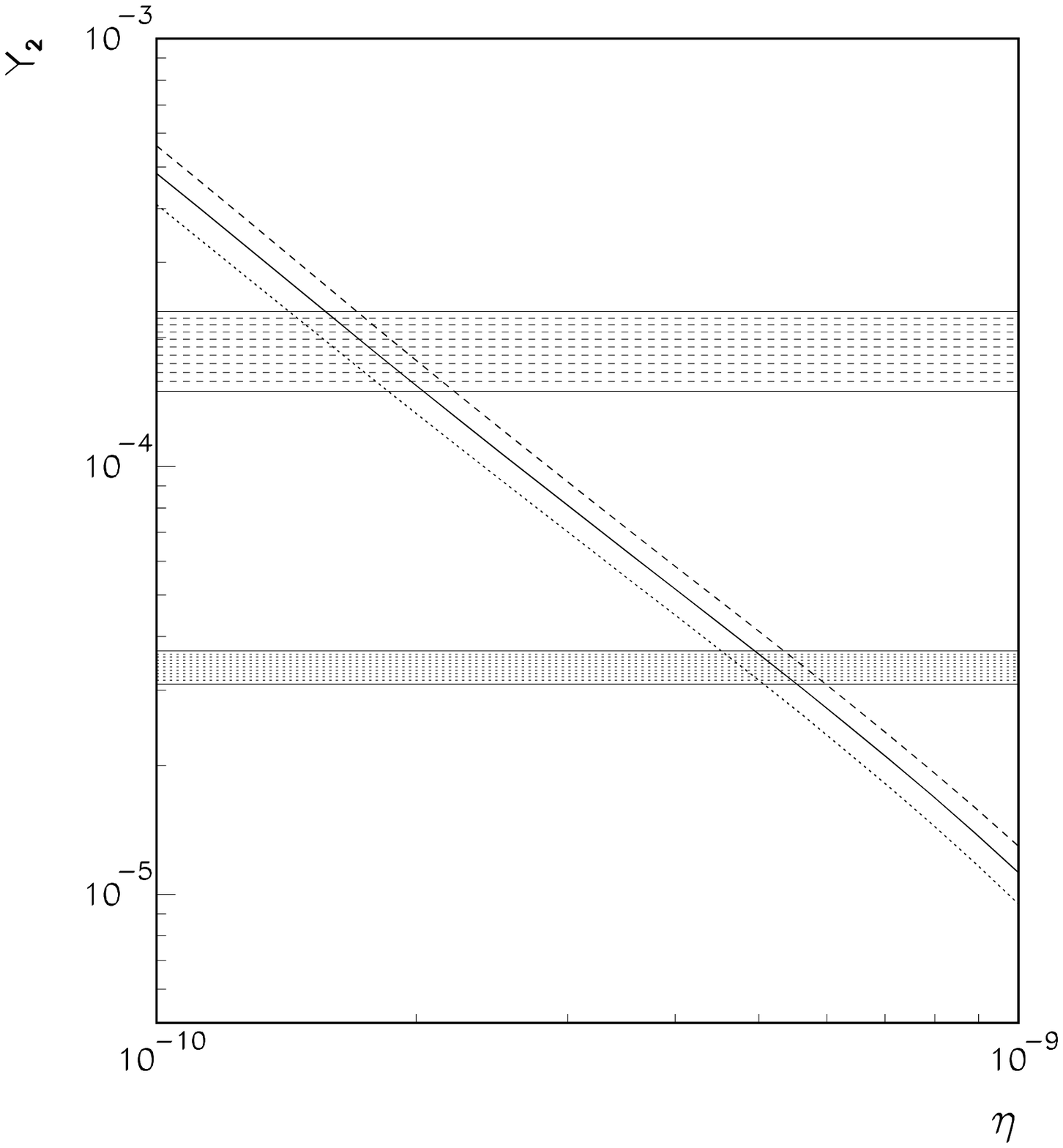}}
\vspace{-1.2truecm}
\caption{The quantity $Y_2$ versus $\eta$ is reported. The same notation of Fig. 3
is used. The horizontal bands dashed and dotted are the experimental values
(see for example \protect\cite{Sarkar}).}
\end{figure}

\begin{figure}[htb]
\vspace{-1cm}
\epsfysize=7.0cm
\epsfxsize=7.0cm
\centerline{\epsffile{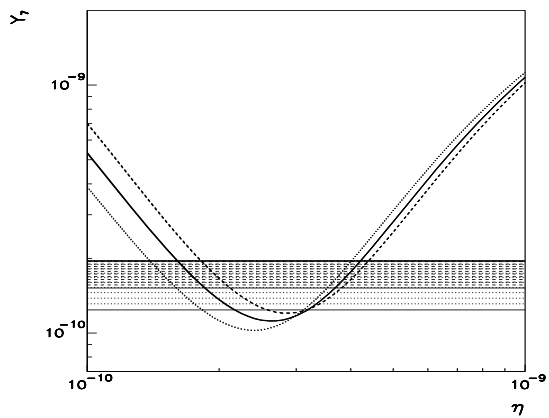}}
\vspace{-1.2truecm}
\caption{The quantity $Y_7$ versus $\eta$. The same notation of Fig. 3
is used. The horizontal bands dashed and dotted are the experimental values
(see for example \protect\cite{Sarkar}).}
\end{figure}

\section{Conclusions}

A detailed study of the effects on primordial abundances of the radiative,
finite nucleon mass, thermal and plasma corrections to Born rates $n
\leftrightarrow p$ has been recently carried out \cite{EMMP2}. This
analysis which has reduced the uncertainty on $Y_4$ to less than $1 \%$ has
been performed using an update version of the BBN standard code
\cite{WagonerKawano}.

\end{document}